\documentclass[twocolumn,showpacs,preprintnumbers,amsmath,amssymb]{revtex4}
 
\usepackage{graphicx}
\usepackage{dcolumn}
\usepackage{bm}
 
\newcommand{\mean}[1]{\langle{#1}\rangle}

\newcommand{\dgg}{^{\dagger}}
\newcommand{\Tr}{{\rm Tr}\hspace{0.07cm}}

\newcommand{\im}{{\rm i}}

\newcommand{\half}{\frac{1}{2}}

\begin{document}
 
\preprint{APS/123-QED}
 
\title{Robust observer for uncertain linear quantum systems
}
 
\author{Naoki Yamamoto}
 \email{naoki@cds.caltech.edu}
\affiliation{%
Physical Measurement and Control 266-33, California 
Institute of Technology, Pasadena, CA 91125
}%
\date{\today}

\begin{abstract}
In the theory of quantum dynamical filtering, one of the biggest issues 
is that the underlying system dynamics represented by a quantum stochastic 
differential equation must be known exactly in order that the corresponding 
filter provides an optimal performance; 
however, this assumption is generally unrealistic. 
Therefore, in this paper, we consider a class of linear quantum systems 
subjected to time-varying norm-bounded parametric uncertainties and then 
propose a robust observer such that the variance of the estimation error is 
guaranteed to be within a certain bound. 
Although in the linear case much of classical control theory can be applied 
to quantum systems, the quantum robust observer obtained in this paper does 
not have a classical analogue due to the system's specific structure with 
respect to the uncertainties. 
Moreover, by considering a typical quantum control problem, we show that 
the proposed robust observer is fairly robust against a parametric 
uncertainty of the system even when the other estimators---the 
optimal Kalman filter and risk-sensitive observer---fail in the estimation. 
\end{abstract}

\maketitle


\section{Introduction}

Quantum filtering theory was pioneered by Belavkin in 
remarkable papers \cite{belavkin1,belavkin2,belavkin3} and was more 
lucidly reconsidered by Bouten {\it et al.} \cite{luc1,luc2}. 
This theory is now recognized as a very important basis for the development 
of various engineering applications of the quantum theory such as quantum 
feedback control \cite{wiseman1,doherty1,thomsen,geremia1,ramon1,wiseman2}, 
quantum dynamical parameter estimation \cite{mabuchi,geremia2,stockton}, 
and quantum information processing \cite{ahn1,ahn2}.

We here provide a brief summary of the quantum filtering theory by using 
the same notations as those in \cite{luc1,luc2}. 
Let us consider an open system in contact with a field, particularly 
a vacuum electromagnetic field. 
This interaction is completely described by a unitary operator $\hat{U}_t$ 
that obeys the following quantum stochastic differential equation (QSDE) 
termed the Hudson-Parthasarathy equation \cite{hudson}: 
\begin{equation}
\label{HP-eq}
   \hbar d\hat{U}_t
      =\Big[ \big(-\im \hat{H}-\half \hat{c}\dgg \hat{c}\big)dt
                        +\hat{c}d\hat{B}_t\dgg
                        -\hat{c}\dgg d\hat{B}_t \Big]\hat{U}_t,~~
   \hat{U}_0=\hat{I}, 
\end{equation}
where $\hat{c}$ and $\hat{H}$ are the system operator and Hamiltonian, 
respectively. 
The quantum Wiener process $\hat{B}_t$, which is a field operator, satisfies 
the following quantum Ito rule: 
\[
   d\hat{B}_t d\hat{B}_t=0,~d\hat{B}_t\dgg d\hat{B}_t=0,~
   d\hat{B}_t d\hat{B}_t\dgg=\hbar dt,~d\hat{B}_t\dgg d\hat{B}_t\dgg=0. 
\]
The time evolution of any system observable $\hat{X}$ under the interaction 
(\ref{HP-eq}) is described by the unitary transformation 
$j_t(\hat{X}):=\hat{U}_t\dgg\hat{X}\hat{U}_t$. 
The infinitesimal change in this transformation is calculated as 
\begin{equation}
\label{qsde}
   \hbar dj_t(\hat{X})
      =j_t({\cal L}\hat{X})dt+j_t([\hat{c}\dgg, \hat{X}])d\hat{B}_t
                             +j_t([\hat{X}, \hat{c}])d\hat{B}_t\dgg. 
\end{equation}
Here, we have defined ${\cal L}\hat{X}:=\im[\hat{H}, \hat{X}]
+\hat{c}\dgg\hat{X}\hat{c}-\half\hat{c}\dgg\hat{c}\hat{X}
-\half\hat{X}\hat{c}\dgg\hat{c}$. 
The field operator after the interaction is determined by 
$\hat{B}_t':=j_t(\hat{B}_t)$. 
In the homodyne detection scheme, we measure the field operator of the 
form $Y_t:=\hat{B}_t'+\hat{B}_t'\mbox{}\dgg$, which results in 
\begin{equation}
\label{output}
   dY_t=j_t(\hat{c}+\hat{c}\dgg)dt+d\hat{B}_t+d\hat{B}_t\dgg. 
\end{equation}
An important fact is that the above observable is {\it self-nondemolition}: 
$[Y_s, Y_t]=0$ for all $s$ and $t$. 
This implies that the observation is a classical stochastic process. 
(For this reason, we omit the ``hat" on $Y_t$, but note that it itself is 
not a $c$-number.) 
It is also noteworthy that $Y_t$ satisfies the {\it quantum nondemolition 
(QND)} condition, $[Y_s, j_t(\hat{X})]=0~\forall s\leq t$, for all system 
observables $\hat{X}$. 
Our goal is to obtain the best estimate of the system observable 
$j_t(\hat{X})$ based on the observations $Y_s~(0\leq s\leq t)$, which 
generate the Von Neumann algebra 
${\cal Y}_t={\rm vN}(Y_s\hspace{0.03cm}:\hspace{0.03cm}0\leq s\leq t)$. 
As in the case of the classical filtering theory, the best estimate in 
the sense of the least mean square error, 
$\mean{(j_t(\hat{X})-X')^2}\rightarrow{\rm min.}$, is given by 
(a version of) the quantum conditional expectation: 
$X'=\pi_t(\hat{X})
:={\mathbb P}(\hspace{0.05cm}j_t(\hat{X})
\hspace{0.05cm}|\hspace{0.05cm}{\cal Y}_t\hspace{0.05cm})$. 
Here, the expectation $\mean{\hat{X}}$ is defined by 
$\mean{\hat{X}}:=\Tr[\hat{X}(\hat{\rho}\otimes\hat{\Phi})]$, where 
$\hat{\rho}$ and $\hat{\Phi}$ represent the system quantum state and the 
field vacuum state, respectively. 
It should be noted that the following two conditions must hold in order 
for the above quantum conditional expectation to be defined: 
First, ${\cal Y}_t$ is a commutative algebra, and second, $j_t(\hat{X})$ 
is included in the commutant of ${\cal Y}_t$. 
But these conditions are actually satisfied as shown above. 
Consequently, the optimal filter for the system dynamics (\ref{qsde}) is 
given by the change in $\pi_t(\hat{X})$ as follows: 
\begin{eqnarray}
& & \hspace*{-1em}
\label{filter-eq}
   \hbar d\pi_t(\hat{X})=\pi_t({\cal L}\hat{X})dt
     +\big[\pi_t(\hat{X}\hat{c}+\hat{c}\dgg\hat{X})
\nonumber \\ & & \hspace*{3em}
     \mbox{}-\pi_t(\hat{X})\pi_t(\hat{c}+\hat{c}\dgg)\big]
                           \big[dY_t-\pi_t(\hat{c}+\hat{c}\dgg)dt\big]. 
\end{eqnarray}
We can further incorporate some control terms into the above equation. 
Typically, a bounded real scalar control input $u_t$, which should be 
a function of the observations $Y_s$ up to time $t$, is included in the 
coefficients of the Hamiltonian. 
We lastly remark that the conditional system state $\hat{\rho}_t$ is 
associated with the system observable by the relation 
$\pi_t(\hat{X})=\Tr(\hat{X}\hat{\rho}_t)$, which leads to the dynamics 
of $\hat{\rho}_t$ termed the stochastic master equation.

A key assumption in the filtering theory is that perfect knowledge about 
the system dynamics model (\ref{qsde}) is required in order 
that the filter (\ref{filter-eq}) provides the best estimate of the 
(controlled) system observable. 
However, this assumption is generally unrealistic, and we depend on only 
an approximate model of the system. 
This not only violates the optimality of the estimation but also 
possibly leads to the instability of the estimation error dynamics. 
This problem is well recognized in the classical filtering theory and 
various alternative estimators for uncertain systems, which are not 
necessarily optimal but robust to the uncertainty, have been proposed. 
(We use the term ``filter" to refer to only the optimal estimator.) 
For example, in a {\it risk-sensitive} control problem in which an 
exponential-of-integral cost function is minimized with respect to the 
control input, it is known that the corresponding risk-sensitive observer 
enjoys enhanced robustness property to a certain type of system uncertainty 
\cite{jacobson,bensou,dupuis}. 
Moreover, by focusing on specific uncertain systems, it is possible 
to design a {\it robust observer} such that the variance of the estimation 
error is guaranteed to be within a certain bound for all admissible 
uncertainties \cite{petersen1,xie,shaked,petersen2}.

It is considered that the above mentioned robust estimation methods are 
very useful in the quantum case since it is difficult to specify the 
exact parameters of a quantum system in any realistic situation: for instance, 
the total spin number of a spin ensemble \cite{stockton}. 
With this background, James has developed a quantum version of the 
risk-sensitive observer for both continuous \cite{james2} and discrete 
cases \cite{james1} and applied it to design an optimal risk-sensitive 
controller for a single-spin system. 
We should remark that, however, the above papers did not provide an example 
of a physical system such that the quantum risk-sensitive observer is 
actually more robust than the nominal optimal filter.

Therefore, in this paper, we focus on the robust observer and develop its 
quantum version. 
More specifically, we consider a class of quantum linear systems subjected 
to time-varying norm-bounded parametric uncertainties and obtain a quantum 
robust observer that guarantees a fixed upper bound on the variance of 
the estimation error. 
Although in the linear case much of classical control theory applies to 
quantum systems, the robust observer obtained in this paper does not have 
a classical analogue in the following sense. 
First, unlike the classical case, the error covariance matrix must be 
symmetrized because of the noncommutativity of the measured system 
observables. 
Second, due to the unitarity of quantum evolution, the uncertainties are 
included in the system representation in a different and more complicated 
way than those in the classical system considered previously; 
as a result, both the structure of the quantum robust observer and the 
proof to derive it differ substantially from those found in 
\cite{petersen1,xie,shaked,petersen2}. 
The other contribution of this paper is that it actually provides a quantum 
system such that both the robust observer and the risk-sensitive observer 
show better performance in the estimation error than the nominal optimal 
filter.

This paper is organized as follows. 
Section II provides a basic description of general linear quantum systems, 
in which case the optimal filter (\ref{filter-eq}) is termed the 
{\it quantum Kalman filter}. 
In addition, we derive a linear risk-sensitive observer. 
In both cases, an explicit form of the optimal control input is provided. 
The quantum version of the robust observer is provided in Section III. 
Section IV discusses robustness properties of the proposed robust 
observer and the risk-sensitive observer by considering a typical quantum 
control problem---feedback cooling of particle motion. 
Section V concludes the paper.

We use the following notations: 
for a matrix $A=(a_{ij})$, the symbols $A^{{\mathsf T}}$ and $A^*$ represent 
its transpose and elementwise complex conjugate of $A$, i.e., 
$A^{\mathsf T}=(a_{ji})$ and $A^*=(a_{ij}^*)=(A\dgg)^{{\mathsf T}}$, 
respectively; 
these rules can be applied to any rectangular matrix including column and 
row vectors. 
A Hermitian matrix $A=A\dgg$ is positive semidefinite if 
$v\dgg Av\geq 0$ for any vector $v$; the inequality $A\geq B$ 
represents the positive semidefiniteness of $A-B$.


\section{Linear quantum system}


\subsection{Quantum Kalman filter}

In this paper, we consider a single one-dimensional particle interacting 
with a vacuum electromagnetic field. 
The extension to the multi-particle case is straightforward 
\cite{wiseman2}. 
In particular, we focus on the particle position $\hat{q}$ and momentum 
$\hat{p}$. 
The system Hamiltonian and operator are respectively given by 
\begin{equation}
\label{linear-case}
    \hat{H}=\half \hat{x}^{{\mathsf T}}G\hat{x}
              -\hat{x}^{{\mathsf T}}\Sigma Bu_t,~~
    \hat{c}=\tilde{C}\hat{x},
\end{equation}
where $\hat{x}=[\hat{q}~\hat{p}]^{{\mathsf T}}$. 
Here, $u_t\in{\mathbb R}$ is the control input, $B\in{\mathbb R}^2$ is a 
column vector, and $\tilde{C}\in{\mathbb C}^2$ is a row vector. 
The $2\times 2$ matrix $G$ is real symmetric and $\Sigma$ is given by 
\[
    \Sigma=
      \left[ \begin{array}{cc}
       0 & 1 \\
       -1 & 0 \\
      \end{array} \right]. 
\]
Then, by defining 
$\hat{x}_t=[\hat{q}_t~\hat{p}_t]^{{\mathsf T}}
=[j_t(\hat{q})~j_t(\hat{p})]^{{\mathsf T}}$ 
and noting the commutation relation $[\hat{q},~\hat{p}]=\im\hbar$, the 
system dynamics (\ref{qsde}) leads to the following linear QSDE: 
\begin{equation}
\label{linear-qsde}
    d\hat{x}_t=A\hat{x}_tdt+Bu_t dt
           +\im\Sigma[\tilde{C}^{{\mathsf T}}d\hat{B}_t\dgg
                     -\tilde{C}\dgg d\hat{B}_t], 
\end{equation}
where the matrix $A$ is defined by 
$A:=\Sigma[G+{\rm Im}(\tilde{C}\dgg\tilde{C})]$. 
The output equation (\ref{output}) becomes 
\[
    dY_t=F\hat{x}_tdt+d\hat{B}_t+d\hat{B}_t\dgg,~~
    F:=\tilde{C}+\tilde{C}^*. 
\]
It follows from Eq. (\ref{filter-eq}) that the best estimate of the system 
observable, 
$\pi_t(\hat{x})
:=[\pi_t(\hat{q})~\pi_t(\hat{p})]^{{\mathsf T}}\in{\mathbb R}^2$, 
obeys the following filter equation: 
\begin{eqnarray}
& & \hspace*{-1em}
\label{linear-filter}
    d\pi_t(\hat{x})=A\pi_t(\hat{x}) dt +Bu_t dt
\nonumber \\ & & \hspace*{1.5em}
    \mbox{}+\Big[\frac{1}{\hbar}V_t F^{{\mathsf T}}
        +\Sigma^{{\mathsf T}}{\rm Im}(\tilde{C})^{{\mathsf T}}\Big]
          (dY_t-F\pi_t(\hat{x}) dt). 
\end{eqnarray}
In the above equation, $V_t$ represents the symmetrized covariance 
matrix defined by 
\begin{eqnarray}
& & \hspace*{-1em}
\label{covariance}
    V_t:={\mathbb P}(\hspace{0.05cm}\hat{P}_t
              \hspace{0.05cm}|\hspace{0.05cm}{\cal Y}_t\hspace{0.05cm})
\nonumber \\ & & \hspace*{-1em}
    \hat{P}_t:=\left[ \begin{array}{cc}
     \Delta\hat{q}_t^2 
      & \half(\Delta\hat{q}_t\Delta\hat{p}_t
             +\Delta\hat{p}_t\Delta\hat{q}_t) \\
     \half(\Delta\hat{q}_t\Delta\hat{p}_t+\Delta\hat{p}_t\Delta\hat{q}_t) 
      & \Delta\hat{p}_t^2 
        \end{array} \right], 
\nonumber \\ & & \hspace*{-1em}
    \mbox{}
\end{eqnarray}
where $\Delta\hat{q}_t:=\hat{q}_t-\pi_t(\hat{q})$ and 
$\Delta\hat{p}_t:=\hat{p}_t-\pi_t(\hat{p})$. 
The covariance matrix $V_t$ changes in time deterministically according 
to the following Riccati differential equation: 
\begin{eqnarray}
& & \hspace*{-1em}
\label{riccati}
    \dot{V}_t=AV_t+V_t A^{{\mathsf T}}+D
\nonumber \\ & & \hspace*{0.5em}
    \mbox{}-\frac{1}{\hbar}
     (V_tF^{{\mathsf T}}
         +\hbar\Sigma^{{\mathsf T}}{\rm Im}(\tilde{C})^{{\mathsf T}})
     (F V_t+\hbar{\rm Im}(\tilde{C})\Sigma), 
\nonumber \\ & & \hspace*{-1em}
    V_0={\mathbb P}(\hspace{0.05cm}\hat{P}_0
               \hspace{0.05cm}|\hspace{0.05cm}{\cal Y}_0\hspace{0.05cm}), 
\end{eqnarray}
where $D:=\hbar\Sigma{\rm Re}(\tilde{C}\dgg\tilde{C})\Sigma^{{\mathsf T}}$. 
Consequently, the optimal filter for the linear quantum system 
(\ref{linear-qsde}) is described by the closed set of equations 
(\ref{linear-filter}) and (\ref{riccati}), which is termed the quantum 
Kalman filter \cite{doherty1,wiseman2,geremia2}. 
A remarkable fact is that the behavior of $V_t$ is determined without 
respect to the output $Y_t$. 
This indicates that we can evaluate the quantum conditional expectation 
$V_t
={\mathbb P}(\hspace{0.05cm}\hat{P}_t
\hspace{0.05cm}|\hspace{0.05cm}{\cal Y}_t\hspace{0.05cm})$ 
by simply calculating the expectation $V_t=\mean{\hat{P}_t}$. 
Actually, as $V_t$ evolves deterministically, we can see 
\[
    \mean{\hat{P}_t}
      =\mean{{\mathbb P}(\hspace{0.05cm}\hat{P}_t
         \hspace{0.05cm}|\hspace{0.05cm}{\cal Y}_t\hspace{0.05cm})}
      =\mean{V_t}=V_t. 
\]

Now, the quantum version of Linear Quadratic Gaussian (LQG) control 
problem is addressed as follows. 
For the {\it linear} quantum system driven by the quantum {\it Gaussian} 
noise, we aim to find an optimal control input $u_t^{{\rm opt}}$, which 
is a function of the observations $Y_s~(0\leq s\leq t)$, such that the 
following {\it quadratic} cost function is minimized: 
\begin{equation}
\label{lqg}
     J[u_t]=
     \Big\langle 
      \int_0^T \Big( \half \hat{x}_t^{{\mathsf T}}M\hat{x}^{\mbox{}}_t
                         +\frac{r}{2}u_t^2 \Big)dt
                         +\half \hat{x}_T^{{\mathsf T}}N\hat{x}_T^{\mbox{}}
       \Big\rangle. 
\end{equation}
The positive semidefinite matrices $M\geq 0,~N\geq 0$ and the scalar 
number $r>0$ reflect our control strategy; 
For example, if we are strongly restricted in the magnitude of the control 
input, a large value of $r$ should be chosen. 
This problem can be solved by using the dynamic programming method. 
The optimal input is then given by 
$u_t^{{\rm opt}}=-(2/r)B^{{\mathsf T}}K_t\pi_t(\hat{x})$, where the real 
symmetric matrix $K_t$ is a solution to the following Riccati differential 
equation: 
\begin{eqnarray}
& & \hspace*{-1em}
\label{lqg-riccati}
   \dot{K}_t+K_t A+A^{{\mathsf T}}K_t
         -\frac{2}{r}K_tBB^{{\mathsf T}}K_t+\half M=O,
\nonumber \\ & & \hspace*{-1em}
   K_T=N. 
\end{eqnarray}
Thus, we observe that the optimal control input $u_t^{{\rm opt}}$ is not a 
function of the entire observation history up to time $t$, but only 
depends on the solution to the Kalman filter (\ref{linear-filter}) and 
(\ref{riccati}) at time $t$. 
A controller that satisfies this desirable property is termed a 
{\it separated controller}. 
A general discussion on the optimality of the separated control is found 
in \cite{luc3}.


\subsection{Quantum risk-sensitive observer}

The risk-sensitive control problem was originally formulated by Jacobson 
within the framework of the classical control theory \cite{jacobson}, 
and recently its quantum version was developed by James 
\cite{james1,james2}. 
The purpose is to design an optimal control input such that the following 
cost function is minimized: 
\begin{equation}
\label{risk-cost}
    J^{\mu}[u_t]=
      \langle \hat{R}_T\dgg j_T^{\mbox{}}({\rm e}^{\mu \hat{\beta}}) 
                         \hat{R}_T^{\mbox{}} \rangle, 
\end{equation}
where $\hat{R}_t$ is the solution to the operator differential equation 
$d\hat{R}_t/dt=(\mu/2)j_t(\hat{\alpha}(u_t))\hat{R}_t$ and the parameter 
$\mu\geq 0$ represents the risk-sensitivity. 
The nonnegative self-adjoint system operators $\hat{\alpha}(u_t)$ and 
$\hat{\beta}$ are termed the running and terminal cost operators, 
respectively. 
In the classical case where the cost operators are scalar values, i.e., 
$\hat{\alpha}(u_t)=\alpha(u_t)$ and $\hat{\beta}=\beta$, the cost function 
(\ref{risk-cost}) is reduced to 
\[
    J^{\mu}[u_t]=
      \Big\langle
         {\rm exp}\Big( \mu 
             \int_0^T \alpha(u_t) dt
                +\mu\beta \Big) \Big\rangle. 
\]
For this reason, Eq. (\ref{risk-cost}) is considered as a natural 
noncommutative generalization of the exponential-of-integral cost function. 
James has proved that the quantity (\ref{risk-cost}) is expressed as 
\begin{equation}
\label{risk-cost-y}
    J^{\mu}[u_t]=
      {\mathbb E}^{\mu}\Big[
         {\rm exp}
            \Big( \mu \int_0^T \pi^{\mu}_t(\hat{\alpha})dt \Big) 
               \pi^{\mu}_T({\rm e}^{\mu\hat{\beta}}) 
                 \Big], 
\end{equation}
where ${\mathbb E}^{\mu}$ denotes the expectation with respect to a 
certain classical probability distribution (see \cite{james2}) and 
$\pi^{\mu}_t(\bullet)$ is a risk-dependent estimate of the system observable. 
The estimator is determined by the following equation: 
\begin{eqnarray}
& & \hspace*{-1em}
\label{risk-observer}
   \hbar d\pi^{\mu}_t(\hat{X})=\pi^{\mu}_t({\cal L}\hat{X})dt
\nonumber \\ & & \hspace*{4em}
   \mbox{}+\frac{\mu}{2}\big[ \pi^{\mu}_t(\hat{X}\hat{\alpha}
           +\hat{\alpha}\hat{X})
        -2\pi^{\mu}_t(\hat{X})\pi^{\mu}_t(\hat{\alpha}) \big]dt
\nonumber \\ & & \hspace*{4em}
   \mbox{}+\big[\pi^{\mu}_t(\hat{X}\hat{c}+\hat{c}\dgg\hat{X})
          -\pi^{\mu}_t(\hat{X})\pi^{\mu}_t(\hat{c}+\hat{c}\dgg)\big]
\nonumber \\ & & \hspace*{6em}
          \times \big[dY_t-\pi^{\mu}_t(\hat{c}+\hat{c}\dgg)dt\big]. 
\end{eqnarray}
This differs from the filtering equation (\ref{filter-eq}) in that 
the risk-dependent term is added to it. 
Therefore, $\pi^{\mu}_t(\hat{X})$ is no longer the optimal estimate of 
the system observable. 
However, the risk-dependent term is indeed necessary in order for the cost 
function (\ref{risk-cost}) to be expressed only in terms of quantities 
defined on the system space that is driven by the output $Y_t$. 
This implies that our knowledge about the system is tempered by purpose.

We now apply the above mentioned risk-sensitive control theory to the 
linear system (\ref{linear-qsde}) with the following cost operators: 
\[
    \hat{\alpha}(u_t)=\half \hat{x}^{{\mathsf T}}M\hat{x}+\frac{r}{2}u_t^2,~~
    \hat{\beta}=\half \hat{x}^{{\mathsf T}}N\hat{x}, 
\]
where $M$ and $N$ are $2\times 2$ positive semidefinite matrices and 
$r>0$. 
Then, through a lengthy calculation we obtain the corresponding observer 
equation as follows: 
\begin{eqnarray}
& & \hspace*{-1.5em}
\label{risk-filter}
    d\pi^{\mu}_t(\hat{x})=(A+\mu V^{\mu}_t M)\pi^{\mu}_t(\hat{x}) dt +Bu_t dt
\nonumber \\ & & \hspace*{0em}
    \mbox{}+\Big[\frac{1}{\hbar}V^{\mu}_t F^{{\mathsf T}}
        +\Sigma^{{\mathsf T}}{\rm Im}(\tilde{C})^{{\mathsf T}}\Big]
          (dY_t-F\pi^{\mu}_t(\hat{x}) dt), 
\end{eqnarray}
where the time evolution of the symmetrized covariance matrix $V^{\mu}_t$ 
is given by 
\begin{eqnarray}
& & \hspace*{-1em}
\label{risk-riccati-1}
    \dot{V}^{\mu}_t=AV^{\mu}_t+V^{\mu}_t A^{{\mathsf T}}+D
\nonumber \\ & & \hspace*{0.5em}
    \mbox{}-\frac{1}{\hbar}
     (V^{\mu}_tF^{{\mathsf T}}
          +\hbar\Sigma^{{\mathsf T}}{\rm Im}(\tilde{C})^{{\mathsf T}})
     (F V^{\mu}_t+\hbar{\rm Im}(\tilde{C})\Sigma), 
\nonumber \\ & & \hspace*{0.5em}
    \mbox{}+\mu(V^{\mu}_t MV^{\mu}_t
       -\frac{\hbar^2}{4}\Sigma^{{\mathsf T}}M\Sigma), 
\nonumber \\ & & \hspace*{-1em}
    V^{\mu}_0={\mathbb P}(\hspace{0.05cm}\hat{P}_0
               \hspace{0.05cm}|\hspace{0.05cm}{\cal Y}_0\hspace{0.05cm}). 
\end{eqnarray}
Consequently, the risk-sensitive observer (\ref{risk-observer}) in 
linear case reduces to the closed set of equations, (\ref{risk-filter}) and 
(\ref{risk-riccati-1}). 
We also see that the cost function (\ref{risk-cost-y}) is calculated as 
\begin{eqnarray}
& & \hspace*{-1em}
    J^{\mu}[u_t]=
      {\mathbb E}^{\mu}\Big[
         {\rm exp}
            \Big( \mu \int_0^T 
               (\half\pi^{\mu}_t(\hat{x})^{{\mathsf T}}M\pi^{\mu}_t(\hat{x})
                   +\frac{r}{2}u_t^2)dt \Big) 
\nonumber \\ & & \hspace*{3em}
    \mbox{}\times
         {\rm exp}
            \Big( \mu \int_0^T 
               \half\Tr(MV_t^{\mu})dt \Big) 
         \pi^{\mu}_T({\rm e}^{\mu\hat{\beta}}) 
                 \Big]. 
\nonumber
\end{eqnarray}
Note that the second integral in the above equation is a constant term as 
$V^{\mu}_t$ is deterministic. 
This is completely a classical controller design problem and was 
already solved by Jacobson \cite{jacobson}; 
the optimal control input that minimizes $J^{\mu}[u_t]$ is given by 
\[
     u_t^{{\rm opt}}=-\frac{2}{r}B^{{\mathsf T}}K^{\mu}_t\pi^{\mu}_t(\hat{x}), 
\]
where $K^{\mu}_t$ satisfies the following Riccati differential equation 
\begin{eqnarray}
& & \hspace*{-1em}
\label{risk-riccati-2}
    \dot{K}^{\mu}_t+K^{\mu}_t A + A^{{\mathsf T}}K^{\mu}_t
             -\frac{2}{r}K^{\mu}_tBB^{{\mathsf T}}K^{\mu}_t+\half M
\nonumber \\ & & \hspace*{-0.8em}
    \mbox{}
     +2\mu K^{\mu}_t
        \Big[\frac{1}{\hbar}V^{\mu}_tF^{{\mathsf T}}
           +\Sigma^{{\mathsf T}}{\rm Im}(\tilde{C})^{{\mathsf T}}\Big]
        \Big[\frac{1}{\hbar}F V^{\mu}_t
           +{\rm Im}(\tilde{C})\Sigma\Big] K^{\mu}_t 
\nonumber \\ & & \hspace*{-0.8em}
    \mbox{}
     +\mu(K^{\mu}_t V^{\mu}_t M + M V^{\mu}_t K^{\mu}_t)=O, 
\nonumber \\ & & \hspace*{-1em}
    K_T^{\mu}=\half\big[
           (I-\mu NV^{\mu}_T)^{-1}N+N(I-\mu V^{\mu}_T N)^{-1} \big]. 
\end{eqnarray}
Therefore, $u_t^{{\rm opt}}$ is a separated controller composed of the 
solutions to the observer equation (\ref{risk-filter}) and the two coupled 
Riccati equations (\ref{risk-riccati-1}) and (\ref{risk-riccati-2}). 
It is notable that these set of equations are identical to those in the 
quantum LQG optimal control problem when the risk parameter $\mu$ 
is zero. 
In this sense, the LQG optimal controller is sometimes referred to as 
the linear {\it risk-neutral} controller.


\section{Robust observer for uncertain linear quantum systems}

This paper deals with a linear quantum system such that specific 
uncertainties are included in the system Hamiltonian $\hat{H}$ and the 
system operator $\hat{c}$ as follows: 
\begin{eqnarray}
& & \hspace*{-1em}
    \hat{H}=\half \hat{x}^{{\mathsf T}}(G+\Delta G_t)\hat{x}
               -\hat{x}^{{\mathsf T}}\Sigma Bu_t,
\nonumber \\ & & \hspace*{-1em}
    \hat{c}=(\tilde{C}+\Delta\tilde{C}_t)\hat{x},
\nonumber
\end{eqnarray}
where the real symmetric matrix $\Delta G_t$ and the complex row vector 
$\Delta\tilde{C}_t$ represent time-varying parametric uncertainties that 
satisfy the following bounds:
\begin{eqnarray}
\label{uncertain-bound-1}
& & \hspace*{-1em}
    (\Delta G_t)^2\leq gI, 
\\ & & \hspace*{-1em}
\label{uncertain-bound-2}
    ({\rm Re}\Delta\tilde{C}_t)^{{\mathsf T}}
          ({\rm Re}\Delta\tilde{C}_t)\leq r_1 I,~~
    ({\rm Im}\Delta\tilde{C}_t)^{{\mathsf T}}
          ({\rm Im}\Delta\tilde{C}_t)\leq r_2 I. 
\nonumber \\ & & \hspace*{-1em}
    \mbox{}
\end{eqnarray}
Here, the nonnegative scalar constants $r_1, r_2$, and $g$ are known 
($I$ denotes the $2\times 2$ identity matrix). 
By defining 
\[
     \Delta A_t:=\Sigma\Delta G_t
        +\Sigma{\rm Im}\Big[
           \tilde{C}\dgg\Delta\tilde{C}_t+\Delta\tilde{C}_t\dgg\tilde{C}
              +\Delta\tilde{C}_t\dgg\Delta\tilde{C}_t \Big], 
\]
the dynamics of the system observable 
$\hat{x}_t=[\hat{q}_t~\hat{p}_t]^{{\mathsf T}}
=[j_t(\hat{q})~j_t(\hat{p})]^{{\mathsf T}}$ is represented as 
\begin{eqnarray}
& & \hspace*{-1em}
\label{uncertain-qsde}
    d\hat{x}_t=(A+\Delta A_t)\hat{x}_tdt+Bu_t dt
\nonumber \\ & & \hspace*{0em}
    \mbox{}
    +\im\Sigma\big[
       (\tilde{C}+\Delta\tilde{C}_t)^{{\mathsf T}}d\hat{B}_t\dgg
          -(\tilde{C}+\Delta\tilde{C}_t)\dgg d\hat{B}_t\big]. 
\end{eqnarray}
Moreover, the uncertainty is also included in the output equation 
(\ref{output}) as follows: 
\begin{eqnarray}
& & \hspace*{-1em}
\label{uncertain-output}
    dY_t=(F+\Delta F_t)\hat{x}_tdt+d\hat{B}_t+d\hat{B}_t\dgg,
\nonumber \\ & & \hspace*{-1em}
    \Delta F_t:=\Delta\tilde{C}_t+\Delta\tilde{C}_t^*. 
\end{eqnarray}
Here, we should remark that the drift and diffusion terms in Eq. 
(\ref{uncertain-qsde}) and the output equation (\ref{uncertain-output}) are 
affected by the common uncertainty $\Delta\tilde{C}_t$. 
This is because the quantum evolution is restricted to satisfy unitarity 
and the system matrices are thus strongly connected with each other. 
This is indeed an intrinsic feature of quantum systems that is not seen 
in general classical systems.

Motivated from the structure of the Kalman filter (\ref{linear-filter}), 
we aim to design a linear observer of the form 
\begin{equation}
\label{notDetermined}
    dx_t=Rx_tdt+Bu_t dt+kdY_t, 
\end{equation}
where $R$ and $k$ are a matrix and a vector to be determined such that the 
variance of the estimation error is guaranteed to be within a certain bound. 
The vector $x_t=[q_t \hspace{0.2cm} p_t]^{{\mathsf T}}\in{\mathbb R}^2$ 
represents the estimate of the system observable $\hat{x}_t$. 
Note that, as in the case of the risk-sensitive observer, $x_t$ is not 
necessarily the optimal estimate of $\hat{x}_t$. 
Furthermore, we here assume that the control input $u_t$ is fixed to a linear 
function of the observer state, $u_t=Lx_t$, where $L$ is a row vector with 
the size $2$. 
Then, an explicit form of $(R,k)$ that enjoys a guaranteed estimation 
error bound is provided in the following theorem. 
We remark again that the theorem can be easily generalized to the 
multi-particle case. 
\\
\\
{\bf Theorem 1.}~
Suppose there exist positive scalars $\delta_i~(i=1,2)$ and 
$\epsilon_i~(i=1,\ldots,8)$ such that the following two coupled Riccati 
equations have positive definite solutions $P_1>0$ and $P_2>0$: 
\begin{eqnarray}
& & \hspace*{-1em}
\label{riccati1}
   (A+BL)P_1+P_1(A+BL)^{{\mathsf T}}+P_1Q_1P_1
\nonumber \\ & & \hspace*{0em}
    \mbox{}
    +D'+\delta_1 I=O, 
\\ & & \hspace*{-1em}
\label{riccati2}
   A'P_2+P_2A'\mbox{}^{{\mathsf T}}+D'+\delta_2 I
\nonumber \\ & & \hspace*{0em}
   \mbox{}
   -\frac{1}{\mu_2}
   (P_2F'\mbox{}^{{\mathsf T}}
       +\mu_1\Sigma^{{\mathsf T}}{\rm Im}(\tilde{C})^{{\mathsf T}})
   (F'P_2+\mu_1{\rm Im}(\tilde{C})\Sigma)
\nonumber \\ & & \hspace*{0em}
   \mbox{}
   -P_2(L^{{\mathsf T}}B^{{\mathsf T}}P_1^{-1}+P_1^{-1}BL)P_2=O, 
\end{eqnarray}
where the matrices $A'$ and $D'$ and the vector $F'$ are defined by 
\begin{eqnarray}
& & \hspace*{-1em}
   A':=A+(D+Q_2+Q_3)P_1^{-1},~~
   D':=D+Q_2+Q_3,
\nonumber \\ & & \hspace*{-1em}
   F':=F+\mu_1{\rm Im}(\tilde{C})\Sigma P_1^{-1}. 
\nonumber
\end{eqnarray}
The definition of the matrices $Q_i~(i=1,2,3)$ are given in Appendix A: 
Eqs. (\ref{Q1}), (\ref{Q2}), and (\ref{Q3}). 
The scalars $\mu_1$ and $\mu_2$ are given by 
$\mu_1=\hbar+4r_1/\epsilon_2$ and 
$\mu_2=\hbar+8r_1/\epsilon_2+\hbar\epsilon_8$, respectively. 
Then, the observer 
\begin{eqnarray}
& & \hspace*{-2em}
\label{robust-filter}
    dx_t=(A'-P_2L^{{\mathsf T}}B^{{\mathsf T}}P_1^{-1})x_t dt
            +Bu_t dt
\nonumber \\ & & \hspace*{-0.2em}
    \mbox{}
      +\frac{1}{\mu_2}\Big[ P_2 F'\mbox{}^{{\mathsf T}}
        +\mu_1\Sigma^{{\mathsf T}}{\rm Im}(\tilde{C})^{{\mathsf T}}\Big]
            (dY_t-F'x_t dt)
\end{eqnarray}
generates the estimate $x_t=[q_t \hspace{0.2cm} p_t]^{{\mathsf T}}$ 
that satisfies 
\begin{equation}
\label{upper-bound}
    \lim_{t\rightarrow\infty}
       \Big\langle (\hat{q}_t-q_t)^2+(\hat{p}_t-p_t)^2 \Big\rangle
          \leq \Tr P_2, 
\end{equation}
for all admissible uncertainties. 
\\
\\
{\bf Proof.}~
We consider the augmented variable 
$\bar{z}_t=[\hat{x}_t~\hat{x}_t-x_t]^{{\mathsf T}}$, 
where $\hat{x}_t$ and $x_t$ satisfy Eqs. (\ref{uncertain-qsde}) 
and (\ref{notDetermined}), respectively. 
Then, $\bar{z}_t$ obeys the following linear QSDE: 
\begin{equation}
\label{augmented}
     d\bar{z}_t=(\bar{A}+\Delta\bar{A}_t)\bar{z}_tdt
                +\bar{b}_\Delta d\hat{B}_t\dgg+\bar{b}_\Delta^*d\hat{B}_t, 
\end{equation}
where 
\begin{eqnarray}
& & \hspace*{-1.2em}
    \bar{A}
     =\left[ \begin{array}{cc}
       A+BL & -BL \\
       A-R-kF & R \\
      \end{array} \right],
    \Delta\bar{A}_t
     =\left[ \begin{array}{cc}
       \Delta A_t & O \\
       \Delta A_t-k\Delta F_t & O \\
      \end{array} \right], 
\nonumber \\ & & \hspace*{-1.2em}
    \bar{b}_\Delta
     =\left[ \begin{array}{c}
       \im\Sigma(\tilde{C}+\Delta\tilde{C}_t)^{{\mathsf T}} \\
       \im\Sigma(\tilde{C}+\Delta\tilde{C}_t)^{{\mathsf T}}-k \\
      \end{array} \right]. 
\nonumber
\end{eqnarray}
Let us now consider the symmetrized covariance matrix of $\bar{z}$; 
$\bar{V}_{nm}=\mean{\bar{z}_n \bar{z}_m+\bar{z}_m \bar{z}_n}/2,~
(n,m=1,\ldots,4)$. 
This satisfies the following generalized uncertainty relation: 
\[
   \mean{\bar{z}_t\bar{z}_t^{{\mathsf T}}}
     =\bar{V}_t+\frac{\im\hbar}{2}\bar{\Sigma}\geq 0,~~
   \bar{\Sigma}:=
            \left[ \begin{array}{cc}
                \Sigma & \Sigma \\
                \Sigma & \Sigma \\
            \end{array} \right]. 
\]
Noting $\hat{B}_t\hat{\Phi}=0$ and the quantum Ito rule 
$d\hat{B}_t d\hat{B}_t\dgg=\hbar dt$, the time evolution of $\bar{V}_t$ 
is calculated as 
\begin{eqnarray}
& & \hspace*{-1.5em}
   \frac{d}{dt}\bar{V}_t
      =(\bar{A}+\Delta\bar{A}_t)\mean{\bar{z}_t\bar{z}_t^{{\mathsf T}}}
       +\mean{\bar{z}_t\bar{z}_t^{{\mathsf T}}}
              (\bar{A}+\Delta\bar{A}_t)^{{\mathsf T}}
       +\hbar\bar{b}_\Delta^*\bar{b}_\Delta^{{\mathsf T}}
\nonumber \\ & & \hspace*{0.8em}
     =(\bar{A}+\Delta\bar{A}_t)
          \Big[\bar{V}_t+\frac{\im\hbar}{2}\bar{\Sigma}\Big]
\nonumber \\ & & \hspace*{2em}
     \mbox{}+\Big[\bar{V}_t+\frac{\im\hbar}{2}\bar{\Sigma}\Big]
          (\bar{A}+\Delta\bar{A}_t)^{{\mathsf T}}
     +\hbar\bar{b}_\Delta^*\bar{b}_\Delta^{{\mathsf T}}
\nonumber \\ & & \hspace*{0.8em}
     =(\bar{A}+\Delta\bar{A}_t)\bar{V}_t
      +\bar{V}_t(\bar{A}+\Delta\bar{A}_t)^{{\mathsf T}}
      +\bar{D}+\Delta\bar{D}_t. 
\nonumber
\end{eqnarray}
The matrices $\bar{D}$ and $\Delta\bar{D}_t$ are given by 
\begin{eqnarray}
& & \hspace*{-1.1em}
    \bar{D}
        =\left[ \begin{array}{cc}
                D & D \\
                D & D \\
         \end{array} \right]
        -\hbar\left[ \begin{array}{cc}
            O & mk^{{\mathsf T}} \\ 
            km^{{\mathsf T}} & 
            km^{{\mathsf T}}+mk^{{\mathsf T}}-kk^{{\mathsf T}} \\
         \end{array} \right], 
\nonumber \\ & & \hspace*{-1.1em}
    \Delta\bar{D}_t
        =\left[ \begin{array}{cc}
                \Delta D_t & \Delta D_t \\
                \Delta D_t & \Delta D_t \\
         \end{array} \right]
        -\hbar\left[ \begin{array}{cc}
            O & \Delta m_t k^{{\mathsf T}} \\
            k\Delta m_t^{{\mathsf T}} & 
            k\Delta m_t^{{\mathsf T}}+\Delta m_t k^{{\mathsf T}} \\
         \end{array} \right], 
\nonumber
\end{eqnarray}
where 
\begin{eqnarray}
& & \hspace*{-1em}
   \Delta D_t:=\hbar\Sigma
        {\rm Re}\Big[
           \tilde{C}\dgg\Delta\tilde{C}_t+\Delta\tilde{C}_t\dgg\tilde{C}
              +\Delta\tilde{C}_t\dgg\Delta\tilde{C}_t 
                  \Big]\Sigma^{{\mathsf T}}, 
\nonumber \\ & & \hspace*{-1em}
    m:=\Sigma^{{\mathsf T}}{\rm Im}(\tilde{C})^{{\mathsf T}},~~
    \Delta m_t:=\Sigma^{{\mathsf T}}{\rm Im}(\Delta\tilde{C}_t)^{{\mathsf T}}.
\nonumber
\end{eqnarray}
Our goal is to design $R$ and $k$ such that the condition 
\begin{eqnarray}
& & \hspace*{-3em}
\label{proof-ineq}
    \exists\bar{X}>0, 
\nonumber \\ & & \hspace*{-3em}
    \mbox{s.t.}~~
    (\bar{A}+\Delta\bar{A}_t)\bar{X}
      +\bar{X}(\bar{A}+\Delta\bar{A}_t)^{{\mathsf T}}
         +\bar{D}+\Delta\bar{D}_t<0 
\end{eqnarray}
is satisfied for all admissible uncertainties; 
in this case, it follows from the lemma shown in Appendix B that the 
relation $\lim_{t\rightarrow\infty}\bar{V}_t\leq\bar{X}$ is satisfied. 
For this purpose, we utilize the following matrix inequalities: 
For all $\bar{X}$ and the uncertain matrices satisfying Eqs. 
(\ref{uncertain-bound-1}) and (\ref{uncertain-bound-2}), we have
\begin{eqnarray}
& & \hspace*{-1em}
\label{uncertain-ineq-1}
    \Delta\bar{A}_t\bar{X}+\bar{X}\Delta\bar{A}_t^{{\mathsf T}}
     \leq \bar{X}\bar{Q}_1\bar{X} + \bar{Q}_2,
\\ & & \hspace*{-1em}
\label{uncertain-ineq-2}
    \Delta\bar{D}_t \leq \bar{Q}_3. 
\end{eqnarray}
The proof of the above inequalities and the definition of the matrices 
$\bar{Q}_i~(i=1,2,3)$ are given in Appendix~A. 
Therefore, the condition (\ref{proof-ineq}) holds for all admissible 
uncertainties if there exists a positive definite matrix $\bar{X}>0$ 
such that the following Riccati inequality holds: 
\[
   \bar{\Psi}
     :=\bar{A}\bar{X}+\bar{X}\bar{A}^{{\mathsf T}}
       +\bar{X}\bar{Q}_1\bar{X}+\bar{D}+\bar{Q}_2+\bar{Q}_3<0. 
\]
Especially we here aim to find a solution of the form 
$\bar{X}={\rm diag}\{P_1, P_2\}$ with $P_1$ and $P_2$ denoting 
$2\times 2$ positive definite matrices. 
Then, partitioning the $4\times 4$ matrix $\bar{\Psi}$ into 
$\bar{\Psi}=(\Psi_{ij})$ with $2\times 2$ matrices $\Psi_{ij}$, 
we obtain 
\begin{eqnarray}
& & \hspace*{-1em}
   \Psi_{11}
     =(A+BL)P_1+P_1(A+BL)^{{\mathsf T}}+P_1 Q_1 P_1+D', 
\nonumber \\ & & \hspace*{-1em}
   \Psi_{21}
     =(A-R-kF)P_1+D'-\mu_1 km^{{\mathsf T}}
      -P_2L^{{\mathsf T}}B^{{\mathsf T}}, 
\nonumber \\ & & \hspace*{-1em}
   \Psi_{22}
     =RP_2+P_2R^{{\mathsf T}}+D'
        +\mu_1(km^{{\mathsf T}}+mk^{{\mathsf T}})
             +\mu_2 kk^{{\mathsf T}}. 
\nonumber
\end{eqnarray}
Let us now assume that the Riccati equation (\ref{riccati1}), which is 
equal to $\Psi_{11}=-\delta_1 I<0$, has a solution $P_1>0$. 
Then, the equality $\Psi_{21}=O$ yields 
$R=A'-kF'-P_2L^{{\mathsf T}}B^{{\mathsf T}}P_1^{-1}$. 
Moreover, $\Psi_{22}$ is then calculated as 
\begin{eqnarray}
& & \hspace*{-1em}
   \Psi_{22}=
    A'P_2+P_2A'\mbox{}^{{\mathsf T}}+D'
\nonumber \\ & & \hspace*{-1em}
    \mbox{}
    +\mu_2
     \Big[k-\frac{1}{\mu_2}P_2 F'\mbox{}^{{\mathsf T}}
             -\frac{\mu_1}{\mu_2}m \Big]
     \Big[k-\frac{1}{\mu_2}P_2 F'\mbox{}^{{\mathsf T}}
             -\frac{\mu_1}{\mu_2}m \Big]^{{\mathsf T}}
\nonumber \\ & & \hspace*{-1em}
    \mbox{}
    -\frac{1}{\mu_2}(P_2F'\mbox{}^{{\mathsf T}}+\mu_1 m)
                    (P_2F'\mbox{}^{{\mathsf T}}+\mu_1 m)^{{\mathsf T}}
\nonumber \\ & & \hspace*{-1em}
    \mbox{}
    -P_2(L^{{\mathsf T}}B^{{\mathsf T}}P_1^{-1}+P_1^{-1}BL)P_2. 
\nonumber
\end{eqnarray}
Hence, the optimal $k$ that minimizes the maximum eigenvalue of $\Psi_{22}$ 
is given by 
\[
   k=\frac{1}{\mu_2}\big(P_2 F'\mbox{}^{{\mathsf T}}
        +\mu_1 m\big)
    =\frac{1}{\mu_2}\Big[P_2 F'\mbox{}^{{\mathsf T}}
        +\mu_1\Sigma^{{\mathsf T}}{\rm Im}(\tilde{C})^{{\mathsf T}}\Big]. 
\]
Then, the existence of a solution $P_2>0$ in Eq. (\ref{riccati2}) 
directly implies $\Psi_{22}=-\delta_2 I<0$. 
As a result, we obtain $\bar{\Psi}={\rm diag}\{-\delta_1 I,-\delta_2 I\}<0$, 
which leads to the objective condition (\ref{proof-ineq}). 
Therefore, according to the lemma in Appendix B, we have 
$\lim_{t\rightarrow\infty}\bar{V}_t\leq\bar{X}$. 
Then, as the third and fourth diagonal elements of the matrix $\bar{V}_t$ 
are respectively given by 
$\bar{V}_{33}=\mean{\bar{z}_3^2}=\mean{(\hat{q}_t-q_t)^2}$ 
and $\bar{V}_{44}=\mean{\bar{z}_4^2}=\mean{(\hat{p}_t-p_t)^2}$, 
we obtain Eq. (\ref{upper-bound}). 
$~\blacksquare$
\\
\\
\indent
The basic idea to determine the form of the quantum robust observer 
(\ref{robust-filter}) is found in several papers that deal with uncertain 
linear classical systems \cite{petersen1,xie,shaked,petersen2}. 
However, the structure of the quantum robust observer differs substantially 
from that of the classical robust observer derived in 
\cite{petersen1,xie,shaked,petersen2}. 
The reason for this is as follows. 
First, unlike the classical case, the covariance matrix $V_t$ of the 
augmented system (\ref{augmented}), which is used to express the 
performance of the robust observer, must be symmetrized in order for 
$V_t$ to be a physical observable. 
Second, the uncertainty $\Delta\tilde{C}_t$ appears both in the drift matrix 
$\Delta A_t$ and the diffusion matrix $\Delta D_t$ in complicated ways; 
this is because, as has been previously mentioned, the system matrices are 
strongly connected with each other due to the unitarity of quantum 
evolution. 
The classical correspondence to the uncertain quantum system 
(\ref{uncertain-qsde}) and (\ref{uncertain-output}) has not been studied. 
For this reason, the resulting robust observer (\ref{robust-filter}) and 
the proof to derive it do not have classical analogues. 
Actually, for standard classical systems whose system matrices can be 
specified independently of one another, the process shown in Appendix A 
is unnecessary.

We now present an important property that the quantum robust observer 
should satisfy: 
When the uncertainties are small or zero, the robust observer should be 
close or identical to the optimal quantum Kalman filter, respectively. 
This natural property is proved as follows. 
\\
\\
{\bf Proposition 2.}~
Consider the case where the uncertainties converge to zero: 
$\Delta G_t\rightarrow 0$ and $\Delta\tilde{C}_t\rightarrow 0$. 
Then, there exist parameters $\delta_i~(i=1,2)$ and 
$\epsilon_i~(i=1,\ldots,8)$ such that the robust observer 
(\ref{robust-filter}) converges to the stationary Kalman filter 
(\ref{linear-filter}) with $V_t$ satisfying the Riccati equation 
$\dot{V}_t=0$ in Eq. (\ref{riccati}). 
\\
\\
{\bf Proof.}~
Let us consider the positive parameters $\epsilon_i~(i=1,\ldots,8)$ 
as follows: 
\begin{eqnarray}
& & \hspace*{-1em}
    \epsilon_1=\sqrt{g},~~
    \epsilon_2=\max\{\sqrt{r_1},\sqrt{r_2}\},~~
    \epsilon_3=\max\{\sqrt{r_1},\sqrt{r_2}\}
\nonumber \\ & & \hspace*{-1em}
    \epsilon_4=r_1,~~
    \epsilon_5=r_2,~~
    \epsilon_6=\sqrt{r_1},~~
    \epsilon_7=\sqrt{r_2},~~
    \epsilon_8=\sqrt{r_2}. 
\nonumber
\end{eqnarray}
In this case, for example, the matrix $Q_1$ is calculated as 
\begin{eqnarray}
& & \hspace*{-1em}
    Q_1=(\sqrt{g}+\max\{\sqrt{r_1},\sqrt{r_2}\}+r_1+r_2)I
\nonumber \\ & & \hspace*{1em}
    \mbox{}
    +\max\{\sqrt{r_1},\sqrt{r_2}\}
       (\tilde{C}_1^{{\mathsf T}}\tilde{C}_1
       +\tilde{C}_2^{{\mathsf T}}\tilde{C}_2), 
\nonumber
\end{eqnarray}
which becomes zero as $g\rightarrow 0, r_1\rightarrow 0$, and 
$r_2\rightarrow 0$. 
Similarly, in these limits, we have 
$Q_2\rightarrow 0, Q_3\rightarrow 0, \mu_1\rightarrow \hbar$, and 
$\mu_2\rightarrow \hbar$. 
Then, since Eq. (\ref{riccati1}) is equivalently written as 
\begin{eqnarray}
& & \hspace*{-1em}
   P_1^{-1}(A+BL)+(A+BL)^{{\mathsf T}}P_1^{-1}+Q_1
\nonumber \\ & & \hspace*{0em}
    \mbox{}
    +P_1^{-1}(D'+\delta_1 I)P_1^{-1}=O, 
\nonumber
\end{eqnarray}
the limit $Q_1\rightarrow 0$ implies that the solution of the above equation 
satisfies $P_1^{-1}\rightarrow 0$. 
We then obtain $A'\rightarrow A,~F'\rightarrow F$, and 
$D'\rightarrow D$. 
Therefore, in this case, Eq. (\ref{riccati2}) with $\delta_2=0$ is identical 
to the Riccati equation $\dot{V}_t=0$ in Eq. (\ref{riccati}). 
The robust observer (\ref{robust-filter}) then converges to the stationary 
Kalman filter (\ref{linear-filter}) with $V_t=P_2$. 
$~\blacksquare$
\\
\\
\indent
The above proposition also states that we can find the parameters 
$\delta_i$ and $\epsilon_i$ such that the robust observer 
(\ref{robust-filter}) approximates the stationary Kalman filter when the 
uncertainties are small, because the solutions of the Riccati equations 
(\ref{riccati1}) and (\ref{riccati2}) are continuous with respect to 
the above parameters.

We lastly remark on the controller design. 
In Theorem 1, we have assumed that the control input is a 
linear function $u_t=Lx_t$. 
This is a reasonable assumption in view of the case of the LQG and 
risk-sensitive optimal controllers. 
Hence, it is significant to study the optimization problems of the vector 
$L$ such that some additional specifications are further achieved. 
For example, $L^{{\rm opt}}$ that minimizes the upper bound of the estimation 
error, $\Tr P_2$, is highly desirable. 
However, it is difficult to solve this problem, since the observer dynamics 
depends on $L$ in a rather complicated manner. 
Therefore, the solution to this problem is beyond the scope of this paper.


\section{Example---Feedback cooling of particle motion}

The main purpose of this section is to show that there actually exists an 
uncertain quantum system such that both the robust observer and the 
risk-sensitive observer perform more effectively than the Kalman filter, 
which is no longer optimum for uncertain systems. 
Moreover, we will carry out a detailed comparison of the above three 
observers by considering each estimation error. 
This is certainly significant from a practical viewpoint.

First, let us describe the system. 
The control objective is to stabilize the particle position $\hat{q}$ 
at the origin by continuous monitoring and control. 
In other words, we aim to achieve $\mean{\pi_t(\hat{q})}=\mean{\hat{q}}=0$ 
with a small error variance. 
The system observable is thus given by 
\[
    \hat{c}=\hat{q},~~~\mbox{i.e.,}~~~\tilde{C}=[1~0]. 
\]
For the Hamiltonian part, 
$\hat{H}=\hat{H}^{{\rm free}}+\hat{H}^{{\rm control}}$, 
we assume the following: 
The control Hamiltonian is proportional to the position operator: 
\begin{equation}
\label{fb-Hamiltonian}
    \hat{H}^{{\rm control}}=-u_t\hat{q},~~~\mbox{i.e.,}~~~
    B=\left[ \begin{array}{c}
         0 \\
         1 \\
       \end{array} \right], 
\end{equation}
where $u_t=Lx_t$ is the input, and the free Hamiltonian is of the form 
$\hat{H}^{{\rm free}}=2\hat{p}^2+V(\hat{q})$, where $V(\hat{q})$ denotes 
the potential energy of the particle. 
In general, the potential energy can assume a complicated structure. 
For example, Doherty {\it et al.} \cite{doherty2} have considered a 
nonlinear feedback control problem of a particle in a double-well potential 
$V(\hat{q})=\hat{q}^4-\hat{q}^2$. 
Since the present paper deals with only linear quantum systems, we approximate 
$V(\hat{q})$ to the second order around the origin and consider a spatially 
local control of the particle. 
In particular, we examine the following two approximated free 
Hamiltonians: 
\[
    \hat{H}_1^{{\rm free}}=2\hat{p}^2-0.05\hat{q}^2,~~
    \hat{H}_2^{{\rm free}}=2\hat{p}^2+0.05\hat{q}^2. 
\]
The former is sometimes referred to as an anti-harmonic oscillator, while 
the latter is a standard harmonic oscillator approximation. 
The system matrices corresponding to $\hat{H}^{{\rm free}}$ are 
respectively given by 
\[
     G_1=\left[ \begin{array}{cc}
           -0.05 & 0 \\
           0     & 2 \\
         \end{array} \right],~~
     G_2=\left[ \begin{array}{cc}
           0.05 & 0 \\
           0    & 2 \\
         \end{array} \right]. 
\]
In the case of the harmonic oscillator Hamiltonian, the system is 
autonomously stable at the origin. 
In contrast, in the case of the anti-harmonic oscillator, the system becomes 
unstable when we do not invoke any control. 
However, it is observed that the control Hamiltonian (\ref{fb-Hamiltonian}) 
with an appropriate control input can stabilize the system. 
An example is the LQG optimal controller with the following 
tuning parameters of the cost function (\ref{lqg}): 
\begin{equation}
\label{example-lqg}
    M=\left[ \begin{array}{cc}
         3 & 0 \\
         0 & 1 \\
      \end{array} \right],~~
    r=\frac{1}{5},~~
    N=\left[ \begin{array}{cc}
         2 & 0 \\
         0 & 0 \\
      \end{array} \right]. 
\end{equation}
Figure I illustrates an estimate of the particle position in both the 
unstable autonomous trajectory and the controlled stable trajectory; 
in the latter case, the control objective $\mean{\pi_t(\hat{q})}=0$ is 
actually satisfied.

\begin{figure}
\includegraphics[scale=0.35]{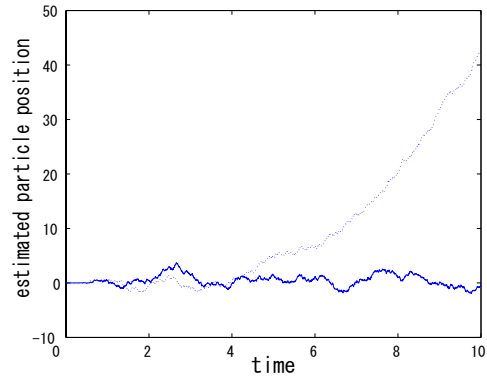}
\caption{\label{fig:comparison}
An example of the unstable autonomous trajectory (dot line) 
and the controlled stable 
trajectory (solid line) shown by $\pi_t(\hat{q})$. 
}
\end{figure}

Second, we describe the uncertainty included in the system. 
In particular, we consider two situations in which uncertain Hamiltonians 
$\Delta\hat{H}_1=-\sqrt{d_t}\hat{q}^2$ and 
$\Delta\hat{H}_2=\sqrt{d_t}\hat{q}^2$ are added to $\hat{H}_1$ and 
$\hat{H}_2$, respectively. 
The unknown time-varying parameter $d_t$ is bounded by the known 
constant $g\geq 0$, i.e., $d_t\in[0, g]$. 
Regarding the uncertainty in the system operator $\hat{c}$, on the other 
hand, we assume $\Delta\tilde{C}_t=0,~\forall t$. 
In this case, we can set $Q_1=\epsilon_1 I, Q_2=(g/\epsilon_1) I$, and 
$Q_3=0$ by choosing the parameters shown in the proof of Proposition 2.

The comparison of the three observers is performed based on the following 
evaluation. 
For the Kalman filter and the risk-sensitive observer, we evaluate the 
stationary mean square error between the ``true" system and the estimator 
for the ``nominal" system corresponding to $d_t=0$ (see Appendix C). 
In both cases, the tuning parameters in the cost function are set to 
Eq. (\ref{example-lqg}). 
Next, for the robust observer, we evaluate the guaranteed upper bound of the 
estimation error $\Tr P_2$ in Eq. (\ref{upper-bound}). 
The control input in the robust observer is set to the stationary LQG 
controller for the nominal system: 
$u_t=Lx_t=-(2/r)B^{{\mathsf T}}K_{\infty}x_t$, where $K_{\infty}$ 
is the stationary solution of Eq. (\ref{lqg-riccati}).

Let us now describe the simulation results. 
First, we consider the case in which the total system Hamiltonian 
is given by 
$\hat{H}=\hat{H}_1^{{\rm free}}+\hat{H}^{{\rm control}}+\Delta\hat{H}_1$. 
Table~I lists the three estimation errors mentioned above for several 
values of $g$. 
Here, the uncertainty $d_t$ is set to the ``worst case" $d_t=g$ for each 
value of $g$. 
In the first row of the table, the notation ``N/A" indicates that the 
solution of the Lyapunov equation (\ref{appendix-lyapunov}) does not 
satisfy $\bar{W}+\im\hbar\bar{\Sigma}/2\geq 0$. 
This implies that the error dynamics between the uncertain actual system 
and the nominal Kalman filter is unstable. 
In other words, the Kalman filter fails in the estimation. 
It should be noted that two excessively large values of the estimation error, 
which appear in the first and second rows, indicate that the error dynamics 
is nearly unstable. 
Therefore, it can be concluded that the Kalman filter and the risk-sensitive 
observer for the nominal system do not work well when the uncertainty 
$d_t~(=g)$ assumes a large. 
On the other hand, as shown in the third row in Table I, the robust observer 
is not very sensitive to the magnitude of the uncertainty and provides a 
good estimation even when $g$ is large. 
The above discussion suggests that the robust observer is possibly the 
best option for dealing with a large uncertainty. 
In other cases, the risk-sensitive observer should be used.

Next, we consider the second example, in which the total system Hamiltonian 
is given by the harmonic oscillator 
$\hat{H}=\hat{H}_2^{{\rm free}}+\hat{H}^{{\rm control}}+\Delta\hat{H}_2$. 
In this case, it is immediately observed in Table II that the estimation 
errors of the robust observer are always greater than those of the others, 
while the risk-sensitive observer shows a good performance, particularly 
when $g$ assumes a large value. 
Hence, in this case the risk-sensitive observer is the most appropriate.

An interesting feature of the robust observer is that in the case of both 
the harmonic and anti-harmonic Hamiltonians, it provides almost the same 
trend in the estimation errors with respect to $g$, whereas the Kalman 
filter and the risk-sensitive observer produce drastically different trends 
in the errors. 
This indicates that the structure of the robust observer is designed 
such that the estimation error is insensitive to the stability property of 
the system. 
However, this design policy sometimes leads to the over conservative 
stability of the error dynamics, and the estimation performance eventually 
reduces.

\begin{table}
\caption{\label{tab:table1}
Comparison of the Kalman, risk-sensitive, and robust observers, denoted by 
``KAL'', ``RSK'', and ``ROB'', respectively. 
The free Hamiltonian of the system is approximated by the anti-harmonic 
oscillator. 
In order to calculate the guaranteed upper bound of the estimation error 
of the robust observer, $\Tr P_2$, parameters $\delta_1$ and $\delta_2$ 
are fixed to $0.1$, and $\epsilon_1$ is selected such that $\Tr P_2$ takes 
the minimum value. 
The risk-sensitive parameter is $\mu=0.3$, and the Planck constant is set 
to unity: $\hbar=1$. 
Note that both the robust observer and the risk-sensitive observer are 
not identical to the Kalman filter even when $g=0$, because the parameters 
$\delta_2$ and $\mu$ are now set to non-zero values. 
}
\begin{ruledtabular}
\begin{tabular}{ccccccc}
$g$ & 0.00 & 0.20 & 0.38 & 0.60 & 0.80 & 0.97 \\
\hline
KAL  & 1.43 & 2.38 & 40.88 & N/A & N/A & N/A \\
RSK & 1.48 & 1.82 & 2.21 & 3.19 & 6.07 & 61.27 \\
ROB  & 1.73 & 3.32 & 4.74 & 7.04 & 10.12 & 14.13 \\
\end{tabular}
\end{ruledtabular}
\end{table}

\begin{table}
\caption{\label{tab:table1}
Comparison of the three types of estimators in the case of the harmonic 
oscillator Hamiltonian. 
All parameters of the estimators are set to the same values in Table I. 
}
\begin{ruledtabular}
\begin{tabular}{ccccccc}
$g$ & 0.00 & 0.20 & 0.40 & 0.60 & 0.80 & 1.00\\
\hline
KAL  & 1.40 & 1.37 & 1.40 & 1.44 & 1.47 & 1.50 \\
RSK & 1.44 & 1.38 & 1.38 & 1.39 & 1.40 & 1.41 \\
ROB  & 1.68 & 3.23 & 4.79 & 6.84 & 9.80 & 14.48 \\
\end{tabular}
\end{ruledtabular}
\end{table}


\section{Conclusion}

In this paper, we have considered a linear quantum system subjected to 
time-varying norm-bounded parametric uncertainties and developed a quantum 
version of the robust observer. 
Although in the linear case much of classical control theory can be 
applied to quantum systems, due to the unitarity of quantum evolution, the 
quantum uncertain system must have a specific structure with respect to 
the uncertainties, and its classical correspondence has not been studied; 
the resulting quantum robust observer has thus no classical analogue. 
The observer differs from both the optimal Kalman filter and the 
risk-sensitive observer; however, it guarantees the upper bound of 
the variance of the estimation error. 
We then investigated the robustness property of the three estimators 
mentioned above by considering a typical quantum control 
problem---feedback cooling of particle motion. 
This examination clarified that the robust observer is superior to the 
others when the autonomous system is unstable and is subjected to an 
unknown perturbation with a large magnitude. 
Therefore, we can conclude that the robust filtering method originally 
developed for classical systems is actually very effective for quantum 
systems as well. 
This fact implies that several robust control techniques in classical 
control theory (e.g., \cite{doyle}) will be applicable to uncertain quantum 
systems.


\begin{acknowledgements}
The author wishes to thank R. van Handel, L. Bouten, and H. Mabuchi for 
their helpful comments. 
This work was supported in part by the Grants-in-Aid for JSPS 
fellows No.06693. 
\end{acknowledgements}


\appendix


\section{Proof of Eqs. (\ref{uncertain-ineq-1}) and (\ref{uncertain-ineq-2})}

At first, we derive a simple yet useful matrix inequality. 
For any real matrices $X$ and $Y$, we obviously have 
\[
     \Big(\sqrt{\epsilon}X-\frac{1}{\sqrt{\epsilon}}Y\Big)^{{\mathsf T}}
     \Big(\sqrt{\epsilon}X-\frac{1}{\sqrt{\epsilon}}Y\Big)\geq O, 
\]
where $\epsilon>0$ is a free parameter. 
The above inequality immediately leads to 
\begin{equation}
\label{basic-ineq}
     X^{{\mathsf T}}Y+Y^{{\mathsf T}}X \leq 
       \epsilon X^{{\mathsf T}}X + \frac{1}{\epsilon}Y^{{\mathsf T}}Y. 
\end{equation}
Next, let us define 
\[
    \tilde{C}_1={\rm Re}\tilde{C},~~
    \tilde{C}_2={\rm Im}\tilde{C},~~
    \Delta\tilde{C}_1={\rm Re}\Delta\tilde{C},~~
    \Delta\tilde{C}_2={\rm Im}\Delta\tilde{C}. 
\]
(In this appendix, we omit the suffix $t$ for simplicity.) 
Then, the conditions (\ref{uncertain-bound-2}) are represented by 
$\Delta\tilde{C}_i^{{\mathsf T}}\Delta\tilde{C}_i\leq r_i I~(i=1,2)$. 
Note that they lead to the scalar inequalities: 
$\Delta\tilde{C}_i\Delta\tilde{C}_i^{{\mathsf T}}\leq r_i~(i=1,2)$.

Now we are at the point to prove. 
Let us first derive the inequality (\ref{uncertain-ineq-1}). 
By a straightforward calculation, we obtain 
\begin{eqnarray}
& & \hspace*{-1em}
   \Delta\bar{A}
    =\bar{\Sigma}\Delta G\bar{E} + \bar{\Theta}_1\Delta\bar{J}_1\bar{E}
     + \bar{\Sigma}\Delta\bar{J}_2\bar{\Theta}_2
\nonumber \\ & & \hspace*{3em}
    \mbox{}
    +\bar{\Sigma}\Delta\tilde{C}_1^{{\mathsf T}}\Delta\tilde{C}_2\bar{E}
    -\bar{\Sigma}\Delta\tilde{C}_2^{{\mathsf T}}\Delta\tilde{C}_1\bar{E}, 
\nonumber
\end{eqnarray}
where 
$\bar{\Sigma}=-[\Sigma~\Sigma]^{{\mathsf T}}\in{\mathbb R}^{4\times 2},~
\bar{E}=[I~O]\in{\mathbb R}^{2\times 4}$ and 
\begin{eqnarray}
& & \hspace*{-1em}
   \bar{\Theta}_1
      =\left[ \begin{array}{cc}
         \Sigma\tilde{C}_1^{{\mathsf T}} 
                          & -\Sigma\tilde{C}_2^{{\mathsf T}} \\
         \Sigma\tilde{C}_1^{{\mathsf T}} 
                          & -\Sigma\tilde{C}_2^{{\mathsf T}}-2k \\
      \end{array} \right],~~
   \bar{\Theta}_2
      =\left[ \begin{array}{cc}
         \tilde{C}_2 & 0^{{\mathsf T}} \\
         -\tilde{C}_1 & 0^{{\mathsf T}} \\
      \end{array} \right],
\nonumber \\ & & \hspace*{-1em}
   \Delta\bar{J}_1
      =\left[ \begin{array}{c}
         \Delta\tilde{C}_2 \\
         \Delta\tilde{C}_1 \\
      \end{array} \right],~~
   \Delta\bar{J}_2
      =[ \Delta\tilde{C}_1^{{\mathsf T}}~~
         \Delta\tilde{C}_2^{{\mathsf T}} ]. 
\nonumber
\end{eqnarray}
We here denoted $0^{{\mathsf T}}=[0~0]$. 
Accordingly, the matrix 
$\Delta\bar{A}\bar{X}+\bar{X}\Delta\bar{A}^{{\mathsf T}}$ is now 
represented by 
\begin{eqnarray}
\label{appendixA-long}
& & \hspace*{-2em}
    \Delta\bar{A}\bar{X}+\bar{X}\Delta\bar{A}^{{\mathsf T}}
\nonumber \\ & & \hspace*{0em}
    =(\bar{\Sigma}\Delta G \bar{E})\bar{X}
       +\bar{X}(\bar{\Sigma}\Delta G \bar{E})^{{\mathsf T}}
\nonumber \\ & & \hspace*{-0.1em}
    \mbox{}
    +(\bar{\Theta}_1\Delta\bar{J}_1 \bar{E})\bar{X}
       +\bar{X}(\bar{\Theta}_1\Delta\bar{J}_1 \bar{E})^{{\mathsf T}}
\nonumber \\ & & \hspace*{-0.1em}
    \mbox{}
    +(\bar{\Sigma}\Delta\bar{J}_2\bar{\Theta}_2)\bar{X}
       +\bar{X}(\bar{\Sigma}\Delta\bar{J}_2\bar{\Theta}_2)^{{\mathsf T}}
\nonumber \\ & & \hspace*{-0.1em}
    \mbox{}
    +(\bar{\Sigma}\Delta\tilde{C}_1^{{\mathsf T}}\Delta\tilde{C}_2
                                                    \bar{E})\bar{X}
    +\bar{X}(\bar{\Sigma}\Delta\tilde{C}_1^{{\mathsf T}}\Delta\tilde{C}_2
                                                    \bar{E})^{{\mathsf T}}
\nonumber \\ & & \hspace*{-0.1em}
    \mbox{}
    -(\bar{\Sigma}\Delta\tilde{C}_2^{{\mathsf T}}\Delta\tilde{C}_1
                                                    \bar{E})\bar{X}
    -\bar{X}(\bar{\Sigma}\Delta\tilde{C}_2^{{\mathsf T}}\Delta\tilde{C}_1
                                                    \bar{E})^{{\mathsf T}}. 
\end{eqnarray}
We are then able to apply Eq. (\ref{basic-ineq}) to evaluate bounds of 
each line in the above equation. 
For example, the second line has the following bound: 
\begin{eqnarray}
& & \hspace*{-1em}
    (\bar{\Sigma}\Delta G \bar{E})\bar{X}
       +\bar{X}(\bar{\Sigma}\Delta G \bar{E})^{{\mathsf T}}
    \leq \epsilon_1\bar{X}\bar{E}^{{\mathsf T}}\bar{E}\bar{X}
        +\frac{1}{\epsilon_1}\bar{\Sigma}\Delta G^2\bar{\Sigma}^{{\mathsf T}}
\nonumber \\ & & \hspace*{10.7em}
    \leq \epsilon_1\bar{X}\bar{E}^{{\mathsf T}}\bar{E}\bar{X}
        +\frac{g}{\epsilon_1}\bar{\Sigma}\bar{\Sigma}^{{\mathsf T}}, 
\nonumber
\end{eqnarray}
where here the assumption on the uncertainty (\ref{uncertain-bound-1}) was 
used. 
The free parameter $\epsilon_1>0$ should be tuned appropriately. 
Next, for evaluating the third line of Eq. (\ref{appendixA-long}), 
we remark the following: 
\begin{equation}
\label{second-line-proof}
    \Delta\bar{J}_1\Delta\bar{J}_1^{{\mathsf T}}
     \leq {\rm diag}\{2r_2,~2r_1\}. 
\end{equation}
This inequality is easily seen; 
the relations 
$\Delta\tilde{C}_i\Delta\tilde{C}_i^{{\mathsf T}}\leq r_i~(i=1,2)$ lead to 
\begin{eqnarray}
& & \hspace*{-1em}
    {\rm det}\Big[
       {\rm diag}\{2r_2,~2r_1\}
       -\Delta\bar{J}_1\Delta\bar{J}_1^{{\mathsf T}} \Big]
\nonumber \\ & & \hspace*{1em}
     =\|\Delta\tilde{C}_1\|^2\|\Delta\tilde{C}_2\|^2
        -\mean{\Delta\tilde{C}_1, \Delta\tilde{C}_2} \geq 0. 
\nonumber
\end{eqnarray}
Here, $\|\bullet\|^2$ and $\mean{\bullet,\bullet}$ denote the standard 
Euclidean norm and inner product, respectively. 
By using Eqs. (\ref{basic-ineq}) and (\ref{second-line-proof}), we then 
obtain the following inequality: 
\begin{eqnarray}
& & \hspace*{-1em}
    (\bar{\Theta}_1\Delta\bar{J}_1 \bar{E})\bar{X}
       +\bar{X}(\bar{\Theta}_1\Delta\bar{J}_1 \bar{E})^{{\mathsf T}}
\nonumber \\ & & \hspace*{1em}
    \leq \epsilon_2\bar{X}\bar{E}^{{\mathsf T}}\bar{E}\bar{X}
        +\frac{1}{\epsilon_2}\bar{\Theta}_1
              \Delta\bar{J}_1\Delta\bar{J}_1^{{\mathsf T}}
                     \bar{\Theta}_1^{{\mathsf T}}
\nonumber \\ & & \hspace*{1em}
    \leq \epsilon_2\bar{X}\bar{E}^{{\mathsf T}}\bar{E}\bar{X}
        +\frac{2}{\epsilon_2}\bar{\Theta}_1
              {\rm diag}\{r_2,~r_1\}\bar{\Theta}_1^{{\mathsf T}}. 
\nonumber
\end{eqnarray}
For the other lines of Eq. (\ref{appendixA-long}), we can use the same 
manner to have their bounds that do not depend on the uncertainties. 
As a result, we obtain the objective inequality 
$\Delta\bar{A}\bar{X}+\bar{X}\Delta\bar{A}^{{\mathsf T}}
\leq \bar{X}\bar{Q}_1\bar{X} + \bar{Q}_2$, where 
\begin{eqnarray}
& & \hspace*{-1em}
    \bar{Q}_1
        :=\left[ \begin{array}{cc}
                Q_1 & O \\
                O & O \\
         \end{array} \right], 
\nonumber \\ & & \hspace*{-1em}
    \bar{Q}_2
        :=\left[ \begin{array}{cc}
                Q_2 & Q_2 \\
                Q_2 & Q_2 \\
         \end{array} \right]
        -\frac{4r_1}{\epsilon_2}
         \left[ \begin{array}{cc}
            O & mk^{{\mathsf T}} \\ 
            km^{{\mathsf T}} & 
            km^{{\mathsf T}}+mk^{{\mathsf T}}-2kk^{{\mathsf T}} \\
         \end{array} \right].
\nonumber
\end{eqnarray}
The matrices $Q_1$ and $Q_2$ are defined as follows. 
\begin{eqnarray}
\label{Q1}
& & \hspace*{-2em}
    Q_1:=(\epsilon_1+\epsilon_2+\epsilon_4+\epsilon_5)I
        +\epsilon_3(\tilde{C}_1^{{\mathsf T}}\tilde{C}_1
               +\tilde{C}_2^{{\mathsf T}}\tilde{C}_2), 
\\ & & \hspace*{-2em}
\label{Q2}
    Q_2:=\Big(\frac{g}{\epsilon_1}+\frac{r_1+r_2}{\epsilon_3}
            +\frac{r_1r_2}{\epsilon_4}+\frac{r_1r_2}{\epsilon_5}\Big)I
\nonumber \\ & & \hspace*{3em}
    \mbox{}
     +\frac{2}{\epsilon_2}
      \Sigma(r_2\tilde{C}_1^{{\mathsf T}}\tilde{C}_1
       +r_1\tilde{C}_2^{{\mathsf T}}\tilde{C}_2)\Sigma^{{\mathsf T}}. 
\end{eqnarray}
The parameters $\epsilon_i>0~(i=1,\ldots,5)$ should be chosen appropriately.

Let us next derive Eq. (\ref{uncertain-ineq-2}). 
Similar to the previous case, we use Eq. (\ref{basic-ineq}) to obtain a 
bound that does not depend on the uncertainty. 
First, we immediately obtain 
\begin{eqnarray}
& & \hspace*{-1em}
    \Delta D=\hbar\Sigma
      \Big[\tilde{C}_1^{{\mathsf T}}\Delta\tilde{C}_1
           +\Delta\tilde{C}_1^{{\mathsf T}}\tilde{C}_1
           +\tilde{C}_2^{{\mathsf T}}\Delta\tilde{C}_2
           +\Delta\tilde{C}_2^{{\mathsf T}}\tilde{C}_2
\nonumber \\ & & \hspace*{5em}
    \mbox{}
    +\Delta\tilde{C}_1^{{\mathsf T}}\Delta\tilde{C}_1
      +\Delta\tilde{C}_2^{{\mathsf T}}\Delta\tilde{C}_2 \Big]
                                           \Sigma^{{\mathsf T}}
\nonumber \\ & & \hspace*{0.9em}
    \leq 
    \hbar\Sigma
      \Big[\epsilon_6\tilde{C}_1^{{\mathsf T}}\tilde{C}_1
           +\frac{r_1}{\epsilon_6}I
           +\epsilon_7\tilde{C}_2^{{\mathsf T}}\tilde{C}_2
           +\frac{r_2}{\epsilon_7}I
\nonumber \\ & & \hspace*{5em}
     \mbox{}+(r_1+r_2)I \Big]\Sigma^{{\mathsf T}}=:Q_3', 
\nonumber
\end{eqnarray}
where $\epsilon_6>0$ and $\epsilon_7>0$ are free parameters. 
This readily leads to 
\[
    \left[ \begin{array}{cc}
       \Delta D & \Delta D \\
       \Delta D & \Delta D \\
    \end{array} \right]
    \leq 
    \left[ \begin{array}{cc}
       Q_3' & Q_3' \\
       Q_3' & Q_3' \\
    \end{array} \right]. 
\]
Also, setting 
$X=[0^{{\mathsf T}}~k^{{\mathsf T}}]$ and 
$Y=[-\Delta m^{{\mathsf T}}~-\Delta m^{{\mathsf T}}]$ in 
Eq. (\ref{basic-ineq}), we obtain the following inequality: 
\begin{eqnarray}
& & \hspace*{-1em}
   -\hbar\left[ \begin{array}{cc}
            O & \Delta m k^{{\mathsf T}} \\
            k\Delta m^{{\mathsf T}} & 
            k\Delta m^{{\mathsf T}}+\Delta m k^{{\mathsf T}} \\
         \end{array} \right]
\nonumber \\ & & \hspace*{1em}
   \leq
   \hbar\left[ \begin{array}{cc}
            \Delta m \Delta m^{{\mathsf T}}/\epsilon_8 & 
               \Delta m \Delta m^{{\mathsf T}}/\epsilon_8 \\ 
            \Delta m \Delta m^{{\mathsf T}}/\epsilon_8 & 
               \Delta m \Delta m^{{\mathsf T}}/\epsilon_8
                  +\epsilon_8 kk^{{\mathsf T}} \\ 
         \end{array} \right]
\nonumber \\ & & \hspace*{1em}
   \leq
   \hbar\left[ \begin{array}{cc}
            r_2I/\epsilon_8 & r_2I/\epsilon_8 \\
            r_2I/\epsilon_8 & r_2I/\epsilon_8+\epsilon_8 kk^{{\mathsf T}} \\ 
         \end{array} \right], 
\nonumber
\end{eqnarray}
where $\epsilon_8>0$. 
Consequently, $\Delta\bar{D}$ is bounded by $\Delta\bar{D}\leq\bar{Q}_3$, 
where
\[
    \bar{Q}_3
        :=\left[ \begin{array}{cc}
                Q_3 & Q_3 \\
                Q_3 & Q_3 \\
         \end{array} \right]
        +\hbar\epsilon_8
         \left[ \begin{array}{cc}
            O & O \\ 
            O & kk^{{\mathsf T}} \\
         \end{array} \right], 
\]
and
\begin{eqnarray}
\label{Q3}
& & \hspace*{-1em}
    Q_3:=\hbar\Big(\frac{r_1}{\epsilon_6}+\frac{r_2}{\epsilon_7}
            +\frac{r_2}{\epsilon_8}+r_1+r_2\Big)I
\nonumber \\ & & \hspace*{3em}
    \mbox{}
     +\hbar
      \Sigma(\epsilon_6\tilde{C}_1^{{\mathsf T}}\tilde{C}_1
       +\epsilon_7\tilde{C}_2^{{\mathsf T}}\tilde{C}_2)\Sigma^{{\mathsf T}}. 
\end{eqnarray}
%


\section{Upper bound lemma}

We consider a matrix-valued differential equation of the form 
\begin{equation}
\label{general-mat-ode}
    \dot{P}_t=AP_t+P_tA^{{\mathsf T}}+BB^{{\mathsf T}}. 
\end{equation}
{\bf Lemma.}~
Suppose there exists a positive definite matrix $X>0$ such that the 
inequality $AX+XA^{{\mathsf T}}+BB^{{\mathsf T}}<0$ holds. 
Then, Eq. (\ref{general-mat-ode}) has a unique stationary solution 
that satisfies 
\[
   \lim_{t\rightarrow\infty}P_t\leq X. 
\]
{\bf Proof.}~
We readily see that the matrix $A$ is strictly stable; any eigenvalue of 
$A$ has a negative real part. 
Now, let us define $\delta P_t:=X-P_t$. 
Then, by using the assumption we have 
\begin{eqnarray}
& & \hspace*{-1em}
    \dot{\delta P}_t
       =-(AX+XA^{{\mathsf T}}+BB^{{\mathsf T}})
        +A\delta P_t+\delta P_tA^{{\mathsf T}}
\nonumber \\ & & \hspace*{0.9em}
    \geq A\delta P_t+\delta P_tA^{{\mathsf T}}, 
\nonumber
\end{eqnarray}
which yields 
$\delta P_t\geq {\rm e}^{At}\delta P_0{\rm e}^{A^{{\mathsf T}}t}$. 
We then obtain $\lim_{t\rightarrow\infty}\delta P_t\geq 0$ since $A$ 
is strictly stable. 
This shows the assertion. 
$~\blacksquare$


\section{Nominal-true systems difference}

The objective here is to characterize the stationary mean square error 
between the ``true" system (\ref{uncertain-qsde}) and the risk-sensitive 
observer (\ref{risk-filter}) specifically designed for the ``nominal" system 
(\ref{linear-qsde}). 
For this purpose, we calculate the symmetrized covariance matrix of the error 
vector $\hat{e}_t:=\hat{x}_t-\pi^{\mu}_t(\hat{x})$, where $\hat{x}_t$ and 
$\pi^{\mu}_t(\hat{x})$ are generated from Eqs. (\ref{uncertain-qsde}) and 
(\ref{risk-filter}), respectively. 
Particularly, we now focus on the stationary observer. 
Thus, let us assume that the two Riccati equations (\ref{risk-riccati-1}) and 
(\ref{risk-riccati-2}) have unique steady solutions $V^{\mu}_{\infty}$ and 
$K^{\mu}_{\infty}$, respectively. 
Then, defining 
\[
    b_o:=\frac{1}{\hbar}V^{\mu}_{\infty}F^{{\mathsf T}}
         +\Sigma^{{\mathsf T}}({\rm Im}\tilde{C})^{{\mathsf T}},~~
    L_o:=-\frac{2}{r}B^{{\mathsf T}}K^{\mu}_{\infty}, 
\]
the stationary risk-sensitive observer is described by 
\begin{eqnarray}
& & \hspace*{-1em}
     d\pi^{\mu}_t(\hat{x})
         =(A+\mu V^{\mu}_{\infty} M+BL_o)\pi^{\mu}_t(\hat{x})dt
\nonumber \\ & & \hspace*{4em}
   \mbox{}+b_o(dY_t-F\pi^{\mu}_t(\hat{x})dt). 
\nonumber
\end{eqnarray}
We then see that the augmented vector 
$\bar{\zeta}_t=[\hat{x}_t~\hat{e}_t]^{{\mathsf T}}$ satisfies 
$d\bar{\zeta}_t=\bar{A}_o\bar{\zeta}_tdt
+\bar{b}_od\hat{B}_t+\bar{b}_o^*d\hat{B}_t\dgg$, where 
\begin{eqnarray}
& & \hspace*{-1em}
    \bar{A}_o
     =\left[ \begin{array}{cc}
       A+\Delta A+BL_o & -BL_o \\
       \Delta A-\mu V^{\mu}_{\infty}M & A+\mu V^{\mu}_{\infty}M-b_oF \\
      \end{array} \right],
\nonumber \\ & & \hspace*{-0.66em}
    \bar{b}_o
     =\left[ \begin{array}{c}
       \im\Sigma\tilde{C}^{{\mathsf T}} \\
       \im\Sigma\tilde{C}^{{\mathsf T}}-b_o \\
      \end{array} \right]. 
\nonumber
\end{eqnarray}
Let $\bar{V}_t$ be the symmetrized covariance matrix of $\bar{\zeta}_t$. 
As mentioned in the proof of Theorem 1, this matrix satisfies 
$\mean{\bar{\zeta}_t\bar{\zeta}_t^{{\mathsf T}}}=
\bar{V}_t+\im\hbar\bar{\Sigma}/2$. 
By using this relation, we obtain 
$d\bar{V}_t/dt=\bar{A}_o\bar{V}_t+\bar{V}_t\bar{A}_o^{{\mathsf T}}+\bar{D}_o$, 
where $\bar{D}_o$ is given by 
\begin{eqnarray}
& & \hspace*{-1em}
    \bar{D}_o
     =\left[ \begin{array}{cc}
                D & D \\
                D & D \\
            \end{array} \right]
\nonumber \\ & & \hspace*{0em}
    -\hbar\left[ \begin{array}{c|c}
       O & [b_o{\rm Im}(\tilde{C})\Sigma]^{{\mathsf T}} \\ \hline
       b_o{\rm Im}(\tilde{C})\Sigma & 
        b_o{\rm Im}(\tilde{C})\Sigma
        +[b_o{\rm Im}(\tilde{C})\Sigma]^{{\mathsf T}}-b_ob_o^{{\mathsf T}} \\
     \end{array} \right]. 
\nonumber
\end{eqnarray}
As a result, the variance of the estimation error is given by 
$\lim_{t\rightarrow\infty}\mean{\hat{e}_t^{\mathsf T}\hat{e}_t}
=\bar{W}_{33}+\bar{W}_{44}$, where $\bar{W}$ is the 
stationary solution of the following Lyapunov equation: 
\begin{equation}
\label{appendix-lyapunov}
     \bar{A}_o\bar{W}+\bar{W}\bar{A}_o^{{\mathsf T}}+\bar{D}_o=O. 
\end{equation}
The estimation error between the true system and the Kalman filter designed 
for the nominal system is immediately evaluated by setting $\mu=0$ in the 
above discussion.


\end{document}